\documentclass[
aps,epsf,
superscriptaddress,
twocolumn,
amsmath,amssymbol,
]
{revtex4-1}

\usepackage[dvipdfmx]{graphicx}

\usepackage{color}

\begin{document}


\title{Spontaneous magnetic field and disorder effects in BaPtAs$_{1-x}$Sb$_x$ with honeycomb network}

\author{Tadashi Adachi}
\email[Corresponding author: ]{t-adachi@sophia.ac.jp}
\affiliation{Department of Engineering and Applied Sciences, Sophia University, 7-1 Kioi-cho, Chiyoda-ku, Tokyo 102-8554, Japan}
\author{Taiki Ogawa}
\affiliation{Department of Physics, Graduate School of Science, Osaka University, 1-1 Machikaneyama, Toyonaka 560-0043, Japan}
\affiliation{Research Institute for Interdisciplinary Science, Okayama University, 3-1-1 Tsushimanaka, Kita-ku, Okayama 700-8530, Japan}
\author{Yota Komiyama}
\affiliation{Department of Engineering and Applied Sciences, Sophia University, 7-1 Kioi-cho, Chiyoda-ku, Tokyo 102-8554, Japan}
\affiliation{Nuclear Structure Research Group, Nishina Center for Accelerator-Based Science, RIKEN, 2-1 Hirosawa, Wako 351-0198, Japan}
\author{Takuya Sumura}
\affiliation{Department of Engineering and Applied Sciences, Sophia University, 7-1 Kioi-cho, Chiyoda-ku, Tokyo 102-8554, Japan}
\author{Yuki Saito-Tsuboi}
\affiliation{Research Institute for Interdisciplinary Science, Okayama University, 3-1-1 Tsushimanaka, Kita-ku, Okayama 700-8530, Japan}
\author{Takaaki Takeuchi}
\affiliation{Research Institute for Interdisciplinary Science, Okayama University, 3-1-1 Tsushimanaka, Kita-ku, Okayama 700-8530, Japan}
\author{Kohei Mano}
\affiliation{Research Institute for Interdisciplinary Science, Okayama University, 3-1-1 Tsushimanaka, Kita-ku, Okayama 700-8530, Japan}
\author{Kaoru Manabe}
\affiliation{Department of Physics, Graduate School of Science, Osaka University, 1-1 Machikaneyama, Toyonaka 560-0043, Japan}
\author{Koki Kawabata}
\affiliation{Department of Engineering and Applied Sciences, Sophia University, 7-1 Kioi-cho, Chiyoda-ku, Tokyo 102-8554, Japan}
\author{Tsuyoshi Imazu}
\affiliation{Department of Mathematics and Physics, Hirosaki University, 3 Bunkyo-cho, Hirosaki 036-8561, Japan}
\author{Akihiro Koda}
\affiliation{Institute of Materials Structure Science, High Energy Accelerator Research Organization (KEK-IMSS), 1-1 Oho, Tsukuba 305-0801, Japan}
\author{Wataru Higemoto}
\affiliation{Advanced Science Research Center, Japan Atomic Energy Agency,2-4 Shirakata, Tokai 319-1195, Japan}
\author{Hirotaka Okabe}
\affiliation{Institute of Materials Structure Science, High Energy Accelerator Research Organization (KEK-IMSS), 1-1 Oho, Tsukuba 305-0801, Japan}
\affiliation{Institute for Materials Research, Tohoku University, 2-1-1 Katahira, Aoba-ku, Sendai 980-8577, Japan}
\author{Jumpei G. Nakamura}
\affiliation{Institute of Materials Structure Science, High Energy Accelerator Research Organization (KEK-IMSS), 1-1 Oho, Tsukuba 305-0801, Japan}
\author{Takashi U. Ito}
\affiliation{Advanced Science Research Center, Japan Atomic Energy Agency,2-4 Shirakata, Tokai 319-1195, Japan}
\author{Ryosuke Kadono}
\affiliation{Institute of Materials Structure Science, High Energy Accelerator Research Organization (KEK-IMSS), 1-1 Oho, Tsukuba 305-0801, Japan}
\author{Christopher Baines}
\affiliation{Laboratory for Muon Spin Spectroscopy, Paul Scherrer Institut, Forschungsstrasse 111, 5232 Villigen PSI, Switzerland}
\author{Isao Watanabe}
\affiliation{Nuclear Structure Research Group, Nishina Center for Accelerator-Based Science, RIKEN, 2-1 Hirosawa, Wako 351-0198, Japan}
\author{Takanori Kida}
\affiliation{Center for Advanced High Magnetic Field Science, Graduate School of Science, Osaka University, 1-1 Machikaneyama, Toyonaka 560-0043, Japan}
\author{Masayuki Hagiwara}
\affiliation{Center for Advanced High Magnetic Field Science, Graduate School of Science, Osaka University, 1-1 Machikaneyama, Toyonaka 560-0043, Japan}
\author{Yoshiki Imai}
\affiliation{Department of Physics, Okayama University of Science, 1-1 Ridai-cho, Kita-ku, Okayama 700-0005, Japan}
\author{Jun Goryo}
\affiliation{Department of Mathematics and Physics, Hirosaki University, 3 Bunkyo-cho, Hirosaki 036-8561, Japan}
\author{Minoru Nohara}
\affiliation{Department of Quantum Matter, Hiroshima University, 1-3-1 Kagamiyama, Higashi-Hiroshima 739-8530, Japan}
\author{Kazutaka Kudo}
\affiliation{Department of Physics, Graduate School of Science, Osaka University, 1-1 Machikaneyama, Toyonaka 560-0043, Japan}
\affiliation{Institute for Open and Transdisciplinary Research Initiatives, Osaka University, 1-1 Yamadaoka, Suita 565-0871, Japan} 
\date{\today}

\begin{abstract}
Chiral superconductivity exhibits the formation of novel electron pairs that breaks the time-reversal symmetry and has been actively studied in various quantum materials in recent years. 
However, despite its potential to provide definitive information, effects of disorder in the crystal structure on the chiral superconductivity has not yet been clarified, and therefore the investigation using a solid-solution system is desirable. 
We report muon-spin-relaxation ($\mu$SR) results of layered pnictide BaPtAs$_{1-x}$Sb$_x$ with a honeycomb network composed of Pt and (As, Sb).
We observed an increase of the zero-field muon-spin relaxation rate in the superconducting (SC) state at the Sb end of $x=1.0$, suggesting the occurrence of spontaneous magnetic field due to the time-reversal symmetry breaking in the SC state. 
On the other hand, spontaneous magnetic field was almost and completely suppressed for the As-Sb mixed samples of $x=0.9$ and 0.2, respectively, suggesting that the time-reversal symmetry breaking SC state in $x=1.0$ is sensitive to disorder. 
The magnetic penetration depth estimated from transverse-field $\mu$SR measurements at $x=1.0$ and 0.2 behaved like weak-coupling $s$-wave superconductivity.
These seemingly incompatible zero-field and transverse-field $\mu$SR results of BaPtAs$_{1-x}$Sb$_x$ with $x=1.0$ could be understood in terms of chiral $d$-wave superconductivity with point nodes on the three-dimensional Fermi surface.

\end{abstract}

\maketitle

\section{Introduction}
Chiral superconductivity is a new class of superconductivity that has attracted much attention over the past decade. 
In a chiral superconducting (SC) state, the orbital motion of SC electron pairs creates a spontaneous magnetic field, which is a manifestation of broken time-reversal symmetry. 
Chiral superconductivity has been proposed in a variety of materials exhibiting a wide variety of pairing symmetries.~\cite{list} 
For example, a chiral singlet state in Sr$_2$RuO$_4$,~\cite{musr,kerr,nmr} $E_{\rm 2u}$ triplet state in UPt$_3$,~\cite{luke-upt3,upt3} chiral $d$-wave state in URu$_2$Si$_2$,~\cite{uru2si2,kittaka} and non-unitary $p$-wave state in Pr$_{1-x}$La$_x$Pt$_4$Ge$_{12}$ \cite{prptge,prlaptge} have been proposed.

A standard way to demonstrate chiral superconductivity is to detect spontaneous magnetic field. 
Muon spin relaxation ($\mu$SR) \cite{textbook} and polar Kerr effect measurements are two techniques that have been used to detect the appearance of spontaneous magnetic field.~\cite{list} 
$\mu$SR is a highly sensitive probe to detect a weak magnetic field in a sample as small as $\sim$1 $\mu$T. 
Muons injected into a sample stop at a potentially stable position in the crystal structure and sense the local magnetic field, causing Larmor precession. 
Muons decay and emit positrons, and the direction of emission has the highest probability of being parallel to the direction of the muon spin at the time of decay.
Therefore, through the spatial distribution of positrons, the information on the local magnetism in the sample is obtained via muon spins. 
$\mu$SR has made significant contributions to the study of strongly correlated electron systems and superconductors.~\cite{adachi-prb69,adachi-prbopt,sonier,koike-adachi}

A honeycomb network has a potential to provide a chiral SC state.
Due to the symmetry of the honeycomb network, the $d_{x^2-y^2}$- and $d_{xy}$-wave symmetries are degenerate in the SC state.
In this case, a ($d_{x^2-y^2} + {\rm i} d_{xy}$)-wave SC symmetry could be realized as the time-reversal symmetry breaking SC state.~\cite{sigrist}
In 2011, Nishikubo {\it et al}. discovered superconductivity with the SC transition temperature $T_{\rm c} \sim 2.4$ K in layered pnictide SrPtAs ($P6_3/mmc$, $D_{6h}^4$, No. 194) with the honeycomb network composed of Pt and As.~\cite{nishikubo} 
Following this discovery, it was theoretically proposed that SrPtAs was a candidate material for chiral $d$-wave superconductivity.~\cite{fischer} 
In fact, Biswas {\it et al}.~\cite{biswas} observed an increase in the muon-spin relaxation rate below $T_{\rm c}$ from zero-field (ZF) $\mu$SR, suggesting the occurrence of spontaneous magnetic field due to the time-reversal symmetry breaking in the SC state. 
Nuclear quadrupole resonance (NQR) \cite{matano} and magnetic penetration depth \cite{landaeta} measurements proposed full-gap spin-singlet superconductivity, while the theory by Ueki {\it et al}.~\cite{ueki} suggested that the chiral $d$-wave state was also able to explain the NQR and penetration-depth results. 
On the other hand, unconventional $f$-wave superconductivity was also proposed from NQR.~\cite{bruckner} 
Therefore, details of the chiral SC state of SrPtAs are not yet clear.

Formerly, we discovered a superconductor with the honeycomb network composed of Pt and (As, Sb), BaPtAs$_{1-x}$Sb$_x$  with the SrPtSb-type structure ($P \bar{6}m2$, $D_{3h}^1$, No. 187).~\cite{kudo,kudo2,ogawa} 
Comparing the Sb end of $x=1.0$ with $x=0.2$, $T_{\rm c}$ of $x=1.0$ is about half of that of $x=0.2$, even though the electronic specific-heat coefficient and the Debye temperature are comparable between $x=1.0$ and 0.2.~\cite{ogawa}
This suggests an unconventional SC state of BaPtAs$_{1-x}$Sb$_x$. 
A theoretical estimation on susceptibilities of the system also supports the unconventional pairing.~\cite{furutani}
In this study, $\mu$SR measurements were carried out in several Sb compositions of BaPtAs$_{1-x}$Sb$_x$ to investigate the possible spontaneous magnetic field, the pairing symmetry of superconductivity and the SC gap.

\section{Experimental}
\begin{figure}[tbp]
\begin{center}
\includegraphics[width=1.0\linewidth]{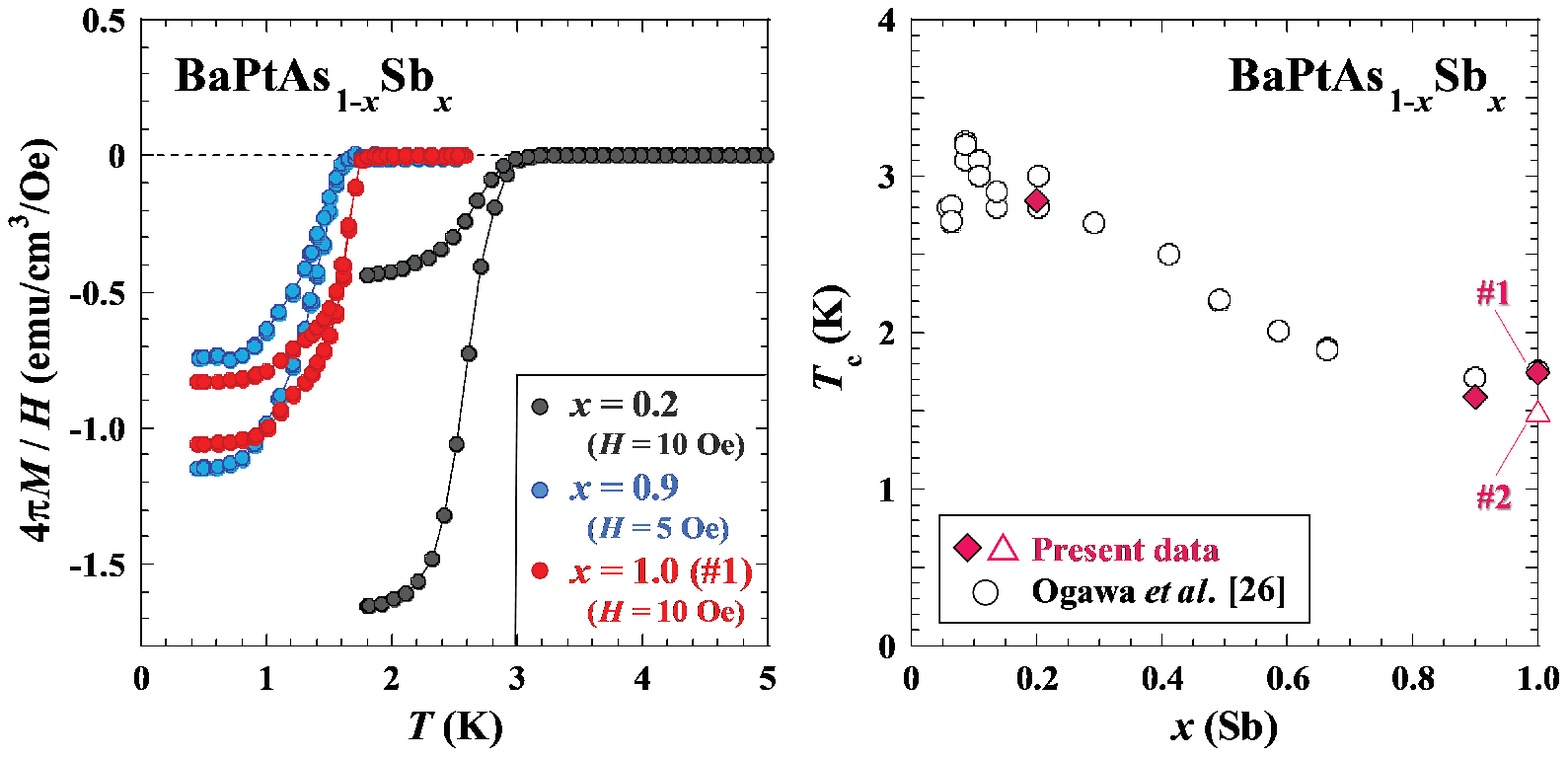}
\caption{(Color online) (left) Temperature dependence of the magnetization of BaPtAs$_{1-x}$Sb$_x$ in magnetic fields of 5 and 10 Oe.  No correction for the demagnetizing field has been made.  (right) Sb-concentration $x$ dependence of $T_{\rm c}$ estimated from the magnetization of BaPtAs$_{1-x}$Sb$_x$.  Open circles are our preceding results from Ref. \cite{ogawa}.}  
\label{fig:fig1}
\end{center}
\end{figure}

Polycrystalline samples of BaPtAs$_{1-x}$Sb$_x$ ($P \bar{6}m2$, $D_{3h}^1$, No. 187) were synthesized by heating a mixture of Ba, Pt, PtAs$_2$, and Sb powders in an alumina crucible sealed in an evacuated quartz tube.
Details of the preparation method have already been published elsewhere.~\cite{kudo,kudo2,ogawa} 
$\mu$SR measurements were performed at $x=0.2, 0.9$ and 1.0(\#1 and \#2). 
Temperature dependence of the magnetization for these samples are shown in Fig. 1(a). 
Magnetization measurements were carried out at low temperatures down to 0.45 K, using a SC quantum interference device magnetometer.
It is found that all samples exhibit bulk superconductivity. 
Figure 1(b) shows the Sb concentration dependence of $T_{\rm c}$ obtained by extrapolating the steepest part of the shielding diamagnetism to zero susceptibility, except for $x=1.0$(\#2) where $T_{\rm c}$ is defined as the temperature of the zero resistivity, together with our preceding results.~\cite{ogawa} 
The sample of $x=0.2$ has high $T_{\rm c}$ of 2.8 K, while $x=0.9$ and 1.0 have low $T_{\rm c}$ values of $1.5-1.7$ K. 
It is noted for $x=1.0$ that $T_{\rm c}$ of \#1 (\#2) is 1.7 (1.5) K. 
Considering the slightly larger normal-state relaxation rate of muon spins for \#2, as described in the results, the quality of the sample is considered to be better for \#1.

To investigate the spontaneous magnetic field, ZF- and longitudinal-field (LF) $\mu$SR measurements were carried out using dilution refrigerator and $^3$He cryostat installed on the D1 and S1 lines of Materials and Life Science Experimental Facility (MLF), J-PARC, Japan, respectively. 
In addition, in order to measure the magnetic penetration depth, transverse-field (TF) $\mu$SR measurements were conducted using dilution refrigerator installed on the LTF spectrometer at Paul Scherrer Institut (PSI), Switzerland and the D1 line at MLF, J-PARC. 
To verify the presence of spontaneous magnetic field in the SC state, it is necessary to reduce the residual magnetic field at the sample position in the $\mu$SR spectrometer as much as possible. 
In the present study, the residual magnetic field was kept as small as possible by using the active field-cancellation technique. 
That is, the magnetic field at the sample position was calculated from the values of the magnetic field measured by magnetic flux meters installed at four locations around the sample position, and current is applied to the compensating coil so that the value approaches zero. 
The magnetic field at the sample position was monitored before and after the measurement using a Hall sensor, and the magnetic field near the surface of the cryostat was monitored during the $\mu$SR measurements. 
As a result, the residual magnetic field during the $\mu$SR measurements could be kept below $\sim 0.7 \mu$T, which is comparable to the value in a previous study in which spontaneous magnetic field was detected.~\cite{feses}

\section{Results}
\begin{figure}[tbp]
\begin{center}
\includegraphics[width=1.0\linewidth]{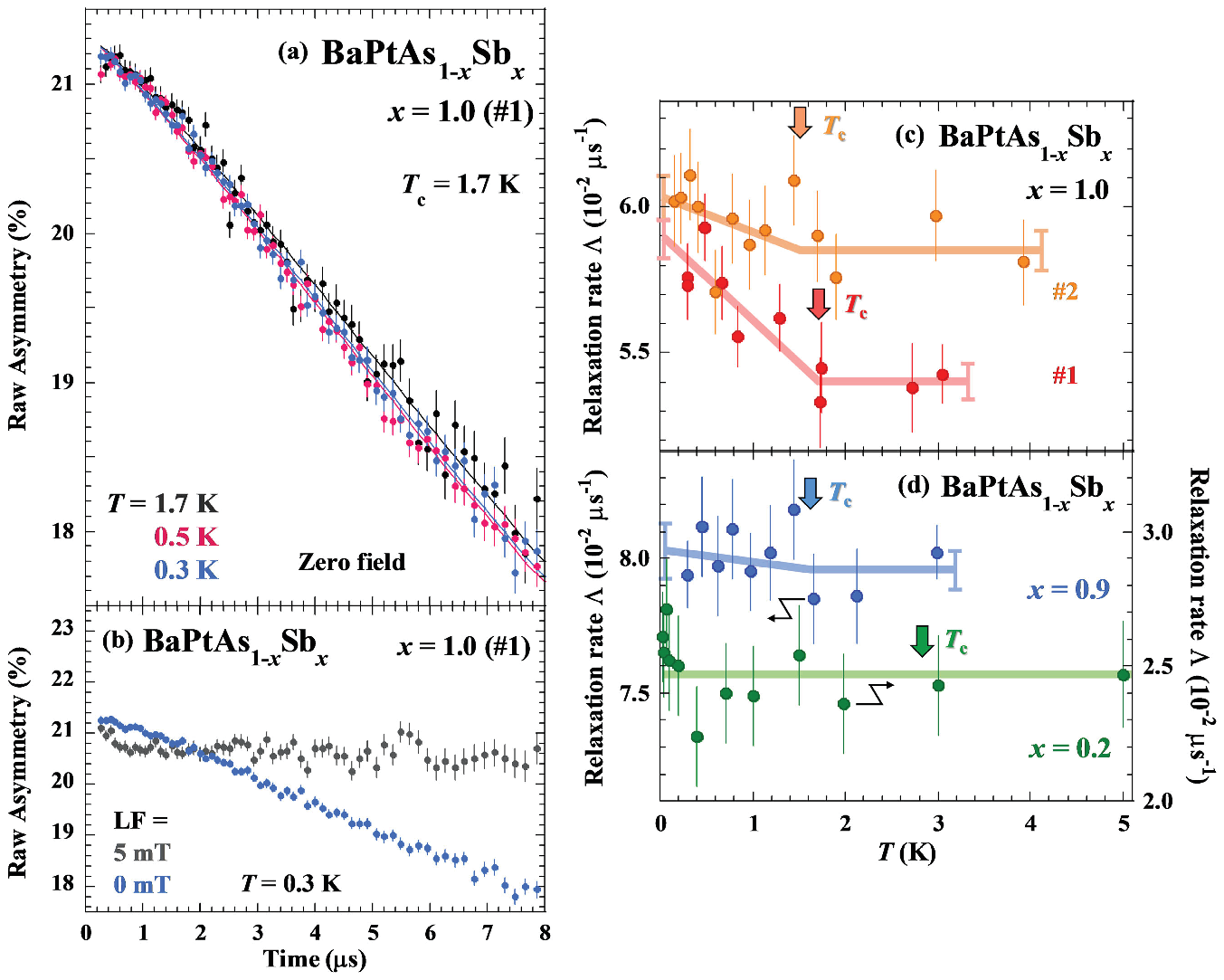}
\caption{(Color online) (a) Zero-field $\mu$SR time spectra below and at $T_{\rm c} = 1.7$ K and (b) longitudinal-field spectra at 0.3 K of BaPtAs$_{1-x}$Sb$_x$ with $x=1.0$(\#1). (c)(d) Temperature dependence of the relaxation rate of muon spins $\Lambda$ in BaPtAs$_{1-x}$Sb$_x$ with $x = 0.2, 0.9, 1.0$(\#1, \#2). Solid lines of $x=0.9$ and 1.0(\#1, \#2) are fitting results assuming that $\Lambda$ is temperature-independent above $T_{\rm c}$ and changes linearly below $T_{\rm c}$. Error bars at the end of fitting lines represent the standard deviation of each fitting.  Solid line of $x=0.2$ is to guide the readers' eye.}  
\label{fig:fig2}
\end{center}
\end{figure}

Figure 2(a) shows typical ZF-$\mu$SR time spectra of BaPtAs$_{1-x}$Sb$_x$ with $x=1.0$(\#1).
The overall behavior of the muon-spin relaxation is Gaussian-like, indicating the dominant contribution of the nuclear spins.
It is found that the relaxation at 0.3 and 0.5 K below $T_{\rm c}$ is slightly but clearly faster than that at $T_{\rm c} = 1.7$ K, suggesting that the internal magnetic field at the muon site increases in the SC state. 
LF-$\mu$SR at LF $= 5$ mT below $T_{\rm c}$ shown in Fig. 2(b) reveals flat spectrum above $\sim 1 \mu$s~\cite{LF}, suggesting that the internal magnetic field has no fluctuating component and therefore it is the quasi-static magnetism.

To obtain detailed information on the internal magnetic field, the spectra were analyzed using the following Equation (1),
\begin{equation}
A(t) = A_0^{\rm ZF} {\rm exp} (-\Lambda t)\, G_{\rm z} (\Delta ,t) + A_{\rm BG}^{\rm ZF}.
\end{equation}
$A_0^{\rm ZF}$ is the initial asymmetry at $t=0$, $\Lambda$ is the relaxation rate of muon spins, $G_{\rm Z}$($\Delta,t$) is the static Kubo-Toyabe function indicating the contribution of the nuclear dipole fields at the muon site.
$\Delta$ is the distribution width of the nuclear dipole fields, $A_{\rm BG}^{\rm ZF}$ is the time-independent background asymmetry due to muons stopped in the Ag plate on which the sample is mounted and in the parts around the sample. 
Equation (1) is the standard one widely used when a Gaussian-like spectrum is observed. 
Since the nuclear spins are regarded as static in the measured temperature range in the $\mu$SR time window, the spectra of the full temperature range were analyzed by the global fit in which $\Delta$ was treated as a common parameter at all temperature.~\cite{delta}
Therefore, the internal magnetic field due to the generation of spontaneous magnetic field is reflected in $\Lambda$.

Figures 2(b) and 2(c) show the temperature dependence of $\Lambda$ of BaPtAs$_{1-x}$Sb$_x$ with $x=0.2, 0.9, 1.0$(\#1, \#2). 
Solid lines of $x=0.9$ and 1.0(\#1, \#2) are fitting results assuming that $\Lambda$ is temperature-independent above $T_{\rm c}$ and changes linearly below $T_{\rm c}$.
Error bars at the end of fitting lines represent the standard deviation of each fitting.
For the Sb end of $x=1.0$, while $\Lambda$ of the \#2 sample increases slightly below $T_{\rm c}$, $\Lambda$ of the \#1 sample exhibits an apparent increase below $T_{\rm c}$. 
The distinct behavior of the \#1 sample is due to the smaller $\Lambda$ values in the normal state above $T_{\rm c}$. 
As $T_{\rm c}$ is also slightly higher, the \#1 sample is considered to be of higher quality of the sample. 
On the other hand, at $x=0.9$, by considering the error of linear fitting, increase in $\Lambda$ is almost absent below $T_{\rm c} = 1.6$ K. 
The $x=0.2$ sample with a high $T_{\rm c}$ value shows no increase in $\Lambda$ below $T_{\rm c} = 2.8$ K. 
These suggest that the spontaneous magnetic field is generated in the SC state in the Sb end of BaPtSb, whereas it almost or completely disappears in the SC state of the As$-$Sb mixed samples.
The magnitude of the increase in $\Lambda$ estimated from the difference of the $\Lambda$ values at $T_{\rm c}$ and at the lowest temperature estimated from the fitting result for $x=1.0$(\#1) and 1.0(\#2) is 0.0040(7) and 0.0015(8) $\mu$s$^{-1}$, corresponding to the magnetic field of $\sim$0.0047(8) mT and $\sim$0.0018(9) mT for \#1 and \#2, respectively. 
The value of \#1 is almost comparable to $\sim$0.007 mT in SrPtAs~\cite{biswas} and is smaller than $\sim$0.05 mT in Sr$_2$RuO$_4$.~\cite{musr}

The magnetic penetration depth $\lambda$ was estimated from TF-$\mu$SR. 
The lower critical field $H_{\rm c1}$ of $x=0.2$ with a high $T_{\rm c}$ value, estimated from the $M$-$H$ curve is 2.8 mT at 2 K. 
Therefore, SC vortices would be uniformly distributed in applied TF = 10 mT. 
The obtained TF-$\mu$SR spectra were fitted using the following Equation (2).
\begin{equation}
\begin{split}
A(t) &= A_0^{\rm TF} {\rm exp}(- \sigma^2 t^2 /2) {\rm cos}(\gamma_{\mu} B_{\rm int}^{\rm TF} t + \phi) \\
& \quad + A_{\rm BG}^{\rm TF}{\rm cos} (\gamma_{\mu} B_{\rm BG}^{\rm TF} t + \phi). \\
\end{split}
\end{equation}
$A_0^{\rm TF}$ is the initial asymmetry at $t=0$, $\sigma$ is the relaxation rate, $B_{\rm int}^{\rm TF}$ ($B_{\rm BG}^{\rm TF}$) is the internal magnetic field at the muon site in (outside) a sample. 
$\gamma_{\mu}$ is the muon gyromagnetic ratio ($\gamma_{\mu} / 2 \pi = 135.5$ MHz/T), $\phi$ is the phase of rotation, $A_{\rm BG}^{\rm TF}$ is the time-independent background asymmetry. 
It is noted that obtained $B_{\rm int}^{\rm TF}$ includes the spontaneous magnetic field in the SC state, which is quite small compared with TF = 10 mT.
Therefore, muons mainly sense the magnetic field composed of the applied TF and its partial expulsion.
The relaxation rate originating from the magnetic field distribution due to the formation of a vortex lattice, $\sigma_{\rm SC}$, is estimated from $\sigma_{\rm SC} = \sqrt{\sigma^2 - \sigma_{\rm nm}^2}$. 
Here, $\sigma_{\rm nm}$ is the relaxation rate due to nuclear spins in the normal state and was taken as the average value of $\sigma$ above $T_{\rm c}$. 
The magnetic penetration depth was then estimated from $\sigma_{\rm SC}$ by the following Equation (3),~\cite{brandt}
\begin{equation}
\frac{2 \sigma_{\rm SC} ^2}{\gamma_\mu^2} = 0.00371 \frac{\Phi_0^2}{\lambda^4}.
\end{equation}
Here $\Phi_0$ is the magnetic flux quantum. 
The temperature dependence of $\lambda$, plotted as $\lambda^{-2} -T$, for BaPtAs$_{1-x}$Sb$_x$ with $x=0.2$ and 1.0(\#2) is shown in Fig. 3(a). 
Both samples exhibit an increase in $\lambda^{-2}$ at low temperatures below $T_{\rm c}$, indicating the formation of a vortex lattice. 
The value of $\lambda^{-2}$ around the lowest temperature is larger for $x=1.0$(\#2). 
Since $\lambda^{-2}$ is proportional to the SC carrier density divided by the effective mass in the clean limit, it is possible that the SC carrier density is larger or the effective mass is smaller at $x=1.0$ than 0.2 where $T_{\rm c}$ is higher. 
The temperature dependence of $B_{\rm int}^{\rm TF}$ for $x=0.2$ and 1.0(\#2) of BaPtAs$_{1-x}$Sb$_x$ is shown in Fig. 3(b). 
Both $B_{\rm int}^{\rm TF}$'s decrease below $T_{\rm c}$, which is due to the expulsion of the magnetic flux by the SC current. 
Although preceding SrPtAs results showed an anomalous increase of $B_{\rm int}^{\rm TF}$ in the SC state,~\cite{biswas} the present results are reasonable.

Since the temperature dependence of $\lambda$ appears to be the behavior of the $s$-wave superconductivity, it was fitted with the following gap function for the isotropic $s$-wave superconductivity,
\begin{equation}
\begin{split}
&\sigma_{\rm SC}[T,\Delta_{\rm SC}(0)] \\
& \quad = 1 + 2 \int_{\Delta_{\rm SC}(T)}^{\infty} \left( \frac{\partial f(E)}{\partial E} \right) \frac{E}{\sqrt{E^2-\Delta_{\rm SC}(T)^2}}dE.\\
\end{split}
\end{equation}
where $f(E)$ is the Fermi distribution function and $\Delta_{\rm SC}$ is the SC gap. 
The temperature dependence of $\Delta_{\rm SC}$ was assumed to be the following equation,
\begin{equation}
\Delta_{\rm SC}(T) = \Delta_{\rm SC}(0) {\rm tanh} \left[ 1.82 \left( 1.018 \left( \frac{T_{\rm c}}{T}-1\right) \right)^{0.51} \right].
\end{equation}
\begin{figure}[tbp]
\begin{center}
\includegraphics[width=0.6\linewidth]{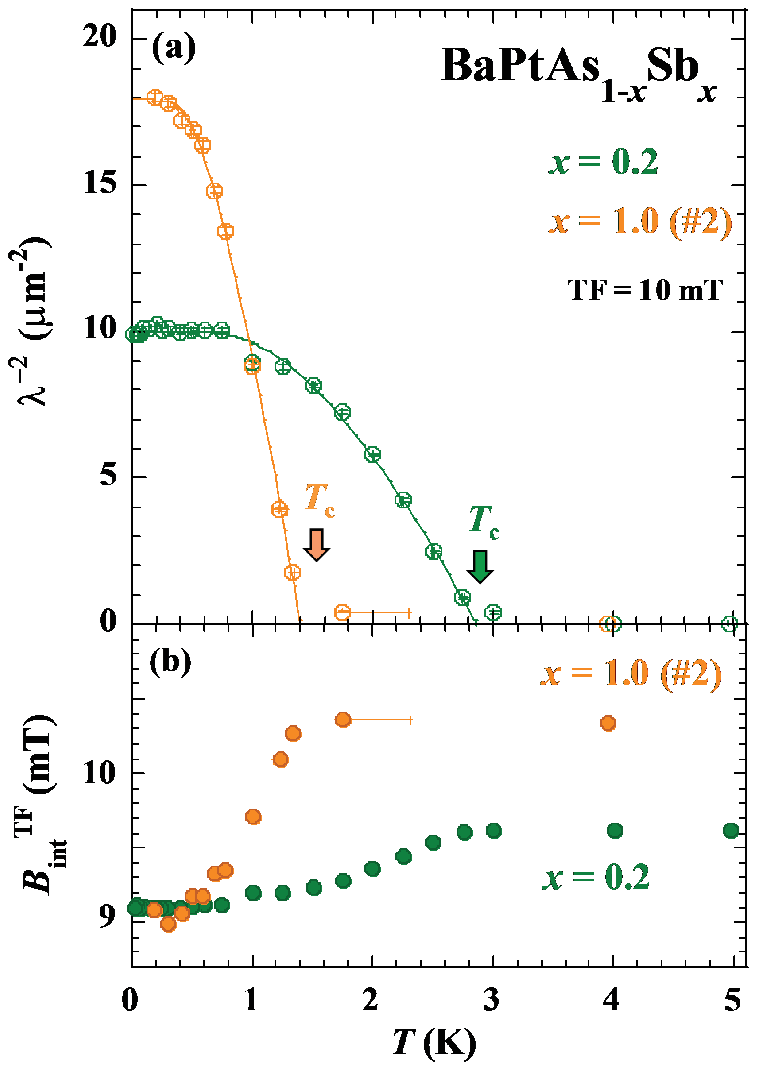}
\caption{(Color online) Temperature dependence of (a) the magnetic penetration depth $\lambda$ plotted as $\lambda^{-2} - T$ and (b) the internal magnetic field at the muon site $B_{\rm int}^{\rm TF}$ in BaPtAs$_{1-x}$Sb$_x$ with $x=0.2, 1.0$(\#2).}  
\label{fig:fig3}
\end{center}
\end{figure}
In Fig. 3(a), the temperature dependence of $\lambda$ of both $x=0.2$ and 1.0(\#2) samples are almost reproduced by Equations (4) and (5). 
The parameters obtained from the fitting are listed in Table I. 
The magnitude of $\Delta_{\rm SC}$ is 0.444(9) and 0.205(3) meV for $x=0.2$ and 1.0(\#2), respectively, which is consistent with the relation of $T_{\rm c}^{\rm TF}$ between samples. 
Calculated $2\Delta_{\rm SC} / {\rm k_B} T_{\rm c}^{\rm TF}$'s are 3.60(7) and 3.41(5) for $x=0.2$ and 1.0(\#2), respectively, which are close to the value of 3.53 for the weak-coupling BCS superconductivity.
This is seemingly incompatible with the time-reversal symmetry breaking superconductivity observed in ZF-$\mu$SR. 
As discussed below, however, it could be understood in terms of chiral $d$-wave superconductivity with point nodes.

\begin{table}[btp]
  \caption{SC parameters of BaPtAs$_{1-x}$Sb$_x$ with $x=0.2$ and 1.0(\#2) obtained from the fitting in Fig. 3(a).}
  \label{table:tab1}
  \centering
  \begin{tabular}{|l|c|c|c|c|}
    \hline
    BaPtAs$_{1-x}$Sb$_x$  & $\lambda(0)$ (nm) &  $\Delta_{\rm SC}$ (meV) & $T_{\rm c}^{\rm TF}$ (K) & $2 \Delta / {\rm k_B} T_{\rm c}^{\rm TF}$ \\
    \hline
    $x=0.2$  & 316.4(2) & 0.449(9) & 2.86(3) & 3.60(7) \\
    \hline
    $x=1.0$(\#2) & 236.4(8) & 0.205(3) & 1.396(9) & 3.41(5) \\
    \hline
  \end{tabular}
\end{table}

\section{Discussion}
From ZF-$\mu$SR of BaPtAs$_{1-x}$Sb$_x$ with the honeycomb network, the increase in the muon-spin relaxation rate $\Lambda$ was observed below $T_{\rm c}$ at $x=1.0$. 
The result of the increase in $\Lambda$ was reproducible, as observed in both the \#1 and \#2 samples, indicating the spontaneous magnetic field in the SC state. 
This suggests a time-reversal symmetry breaking SC state in $x=1.0$. 
On the other hand, the increase in $\Lambda$ below $T_{\rm c}$ was almost and completely absent for $x=0.9$ and 0.2, respectively, where As was partially substituted for Sb. 
This suggests that the spontaneous magnetic field disappears due to the introduction of disorder into the crystal structure by the As substitution. 
The temperature dependence of the magnetic penetration depth estimated from TF-$\mu$SR at $x=0.2$ and 1.0(\#2) of BaPtAs$_{1-x}$Sb$_x$ behaved like weak-coupling $s$-wave superconductivity and deduced $2\Delta_{\rm SC} / {\rm k_B} T_{\rm c}^{\rm TF}$ values were compatible with it. 
This is consistent with formerly reported $\Delta C / \gamma T_{\rm c}$, the specific-heat jump $\Delta C$ divided by the electronic specific-heat coefficient $\gamma$ and $T_{\rm c}$.~\cite{ogawa}
Although the generation of spontaneous magnetic field in the SC state of $x=1.0$(\#2) is small, the TF-$\mu$SR results appear to contradict the time-reversal symmetry breaking at $x=1.0$(\#2). 
These contradictory results are able to be understood as follows. 

A possible candidate of the time-reversal symmetry breaking SC state at $x=1.0$ is the chiral $d$-wave state, a time-reversal symmetry-breaking topological SC state that a system with a hexagonal crystal structure tends to favor.~\cite{fischer,graphene} 
The absence of the spontaneous magnetic field at $x = 0.2$ and 0.9 suggests the change of pairing symmetry, that is, the introduction of disorder into the crystal structure seems like the introduction of non-magnetic impurities significantly suppressing an unconventional pairing like the chiral $d$-wave state.~\cite{sigrist} 
It is noted that the SC gap on a three-dimensional (quasi-two-dimensional) Fermi surface has point nodes (no node).
The temperature dependence of the magnetic penetration depth showing the typical thermal-activation-type behavior at low temperatures,~\cite{ueki} does not necessarily exclude the chiral $d$-wave state, since the power-law behavior from the point-nodal excitation on the three-dimensional Fermi surface could be smeared out when the root mean squares of the Fermi velocities are sufficiently smaller than that of the quasi-two-dimensional Fermi surfaces with full gap excitations.~\cite{ueki} 
Moreover, there are three-dimensional Fermi surfaces close to the ${\rm M}-{\rm L}$ line of the Brillouin zone tracing the saddle points of an energy band.~\cite{uzunok} 
This structure would support the chiral $d$-wave state, whose gap amplitude becomes maximum on the ${\rm M}-{\rm L}$ line.~\cite{fischer,graphene} 
Therefore, the SC symmetry of $x = 1.0$ could not be $s$-wave but chiral $d$-wave with point nodes, which would solve the discrepancy between the ZF and TF results.
Details will be discussed elsewhere.~\cite{imazu}

\section{Summary}

$\mu$SR measurements of the layered pnictide BaPtAs$_{1-x}$Sb$_x$ with the honeycomb network revealed an increase in the zero-field muon-spin relaxation rate below $T_{\rm c}$ at the Sb end of $x=1.0$. 
On the other hand, the increase in the relaxation rate below $T_{\rm c}$ was almost (completely) absent for slightly (heavily) As-substituted $x=0.9$ (0.2). 
From these results, it is highly likely that the spontaneous magnetic field is generated in the SC state at $x=1.0$, that is, a time-reversal symmetry-breaking SC state is realized. 
Although the magnetic penetration depth behaved like weak-coupling $s$-wave superconductivity, the chiral $d$-wave SC state is a possible candidate to explain the present results including the disorder effect and the temperature dependence of the magnetic penetration depth.
Experiments such as angle-dependent thermodynamic measurements~\cite{kittaka} and uniaxial-pressure effects to investigate degenerated pairing symmetries~\cite{klauss} would be helpful to further understand the time-reversal symmetry breaking SC state of BaPtAs$_{1-x}$Sb$_x$.

\section*{Acknowledgments}
We would like to thank H. Kuroe for his help in the magnetization measurements and D. P. Sari for her help in the analysis of TF-$\mu$SR results. 
Part of this work was performed at the Center for Advanced High Magnetic Field Science in Osaka University under the Visiting Researcher's Program of the Institute for Solid State Physics, the University of Tokyo, and at Analytical Instrument Facility, Graduate School of Science, Osaka University, using research equipment shared under the MEXT Project for promoting public utilization of advanced research infrastructure (JPMXS0441200022). 
This work was partly supported by JSPS KAKENHI Grant Number JP19H05823, JP20K03826, JP22H01182, JP23K22453, and JP24K21531 and by JST SPRING, Grant Number JPMJSP2152.



\begin{thebibliography}{99}
\bibitem{LF}
In a short-time region, the relaxation of the LF spectrum is faster than that of the ZF spectrum.  This is probably due to inhomogeneous distribution of vortices caused by vortex pinning.
\bibitem{delta}
Obtained values of $\Delta$ used for the analysis for each sample are 0.0712(18) for $x=1.0$(\#1), 0.0490(24) for $x=1.0$(\#2), 0.0758(19) for $x=0.9$, 0.0987(14) for $x=0.2$.
\end{thebibliography}
\end{document}